\documentclass{article}

\begin{document}

\centerline{\large\bf  Fixed-point Quantum Search for Different Phase
Shifts
\footnote{The paper was supported by NSFC(Grant No. 60433050), the basic research fund
of Tsinghua university No: JC2003043 and partially by the state key lab. of
intelligence technology and system}}
\centerline{Dafa Li$^{a}$\footnote{email address:dli@math.tsinghua.edu.cn},
Xiangrong Li$^{b}$, Hongtao Huang$^{c}$, Xinxin Li$^{d}$ }

\centerline{$^a$ Dept of mathematical sciences, Tsinghua University, Beijing
100084 CHINA}

\centerline{$^b$ Department of Mathematics, University of California, Irvine, CA
92697-3875, USA}

\centerline{$^c$ Electrical Engineering and Computer Science Department} %
\centerline{ University of Michigan, Ann Arbor, MI 48109, USA}

\centerline{$^d$ Dept. of computer science, Wayne State University, Detroit, MI 48202,
USA}


Abstract

Grover recently presented the fixed-point search algorithm. In this letter,
we study the fixed-point search algorithm obtained by replacing equal phase
shifts of $\pi /3$ by different phase shifts.



\section{Introduction}

Grover's search algorithm and fixed-point search algorithm are both used to
find a desired state from an unsorted database. His original algorithm
consists of inversion of the amplitude in the desired state and
inversion-about-average operation\cite{Grover97}. In \cite{Grover98}, Grover
presented a general algorithm: $Q=-I_{\gamma }U^{-1}I_{\tau }U$, where $U$
is any unitary operation, $U^{-1}$ is the adjoint of $U$, $I_{\gamma
}=I-2|\gamma \rangle \langle \gamma |$, $I_{\tau }=I-2|\tau \rangle \langle
\tau |$, $|\gamma \rangle $ is an initial state and $|\tau \rangle $ is a
desired state. When $U^{-1}=U=W$, where $W$ is the Walsh-Hadamard
transformation, and $|\gamma =|0\rangle $, the general algorithm becomes the
original algorithm. Grover showed that the desired state can be found with
certainty after $O(\sqrt{N})$ applications of $Q$ to the initial state $%
|\gamma \rangle $. Long extended Grover's algorithm\cite{Long}. Long's
algorithm is $Q=-I_{\gamma }U^{-1}I_{\tau }U$, where $I_{\gamma
}=I-(-e^{i\theta }+1)|\gamma \rangle \langle \gamma |$, $I_{\tau
}=I-(-e^{i\phi }+1)|\tau \rangle \langle \tau |$. When $\theta =\phi =\pi $,
Long's algorithm becomes Grover's general algorithm. Li et al. proposed that
$U^{-1}$ in Long's algorithm can be replaced by any unitary operation $V$%
\cite{LDF01}\cite{LDF}. In \cite{Galindo}, Galindo et al. gave a family of
Grover's quantum searching algorithms. In \cite{Pang}, Pang et al.
generalized Grover's algorithm and applied their algorithm to image
compression.

Grover's quantum search algorithm can be considered as a rotation of the
state vectors in two-dimensional Hilbert space generated by the initial ($s$%
) and target ($t$) vectors\cite{Grover98}. The amplitude of the target state
increases monotonically towards its maximum and decreases monotonically\
after reaching the maximum \cite{LDF}. As mentioned in \cite{Grover05},
unless we stop as soon as it reaches the target state, it will drift away.
However, the number of iteration steps to find the target state is not an
integer\cite{LDF}. It means that we either stop near the target state or we
drift away.

Grover presented the new algorithm by replacing the selective inversions by
selective phase shifts of $\pi /3$ \cite{Grover05}. The new algorithm
converges to the target state irrespective of the number of iterations. In %
\cite{Tulsi}, an algorithm for obtaining fixed points in iterative quantum
transformations was given and it was illustrated that the algorithm has much
better average-case behavior. In \cite{Boyer}, Boyer et al. described an
algorithm that succeeds with probability approaching to 1.

The quantum search algorithm with different phase rotation angles were
discussed in \cite{Long} \cite{LDF01} \cite{LDF} \cite{Hoyer} \cite{Bilham}.
In this letter, we study the fixed-point search algorithm obtained by
replacing equal phase shifts of $\pi /3$ by different phase shifts and show
that the deviation for different phase shifts is smaller than for equal
phase shifts. However, the smallest average deviation does not occur at
different phase shifts. For the definition of the average deviation, see
section 4.1 of this letter.\ It is well known that the smaller the deviation
is, the more rapidly the algorithm converges to the desired item. Let $%
U_{ts} $ be the amplitude of reaching the target state $|t\rangle $ by
applying $U$ to the start state $|s\rangle $ and $\parallel U_{ts}\parallel
^{2}=1-\epsilon $. For any range $(\beta ,\alpha )$ of $\epsilon $, we argue
that $\alpha $ and $\beta $ determine the phase shifts at which the smallest
average deviation occurs.

This paper is organized as follows. In section 2, we introduce the
fixed-point search algorithm with different phase shifts and derive the
deviation. Section 3 is used to find different phase shifts for small
deviation. Section 3 is devoted to study the smallest average deviation.

\section{The algorithm with different phase shifts}

\subsection{Replacing equal phase shifts of $\protect\pi /3$ by different
phase shifts}

Grover presented the fixed-point search algorithm by replacing the selective
inversions by\ selective phase shifts of $\pi /3$\cite{Grover05}. He
described the transformation $UR_{s}U^{+}R_{t}U$ applied to the start state $%
|s\rangle $, where

\begin{eqnarray}
&&R_{s}=I-(1-e^{i\pi /3})|s\rangle \langle s|,  \nonumber \\
&&R_{t}=I-(1-e^{i\pi /3})|t\rangle \langle t|,  \label{Grove-alg}
\end{eqnarray}

\noindent $t$ stands for the target state.

We consider the transformation $UR_{s}U^{+}R_{t}U$ applied to the start
state $|s\rangle $, where

\begin{eqnarray}
&&R_{s}=I-(1-e^{i\varphi })|s\rangle \langle s|,  \nonumber \\
&&R_{t}=I-(1-e^{i\theta })|t\rangle \langle t|.  \label{diff-shifts}
\end{eqnarray}

\noindent $\theta $ and $\varphi $ are called selective phase shifts. It is
enough to consider $\theta $ and $\varphi $\ to be in $[0,\pi ]$. When $%
\theta =\varphi =\pi /3$, one recovers Grover's fixed-point search algorithm
above. In \cite{LDF05}, the fixed-point search algorithm with two equal
phase shifts was discussed.

We compute $UR_{s}U^{+}R_{t}U|s\rangle =(e^{i\varphi }+(1-e^{i\varphi
})(1-e^{i\theta })\parallel U_{ts}\parallel ^{2})U|s\rangle -(1-e^{i\theta
})U_{ts}|t\rangle $. By the definition of the deviation in \cite{Grover05},
let $D(\theta ,\varphi )$ be the deviation of this superposition from the
state $|t\rangle $. Then we can derive

\begin{eqnarray}
D(\theta ,\varphi )=(1-\parallel U_{ts}\parallel ^{2})\parallel e^{i\varphi
}+(1-e^{i\varphi })(1-e^{i\theta })\parallel U_{ts}\parallel ^{2}\parallel
^{2}.  \label{Dev1}
\end{eqnarray}

\subsection{Simplifying the deviation for different phase shifts}

Grover let $\left| \left| U_{ts}\right| \right| ^{2}=1-\epsilon $, where $%
0<\epsilon <1$. As indicated in \cite{Grover98}, $||U_{ts}||$ is very small
and almost $1/\sqrt{N}$, where $N$ is the size of the database. For example,
$\epsilon =1-1/N>\frac{3}{4}$ when $N>4$. Therefore $\epsilon $ is close to $%
1.$

Substituting $\parallel U_{ts}\parallel ^{2}=1-\epsilon $\ and reducing\ the
second term of $D(\theta ,\varphi )$ in (\ref{Dev1}), we obtain

\begin{eqnarray}
&&D(\theta ,\varphi )=  \nonumber \\
&&\epsilon (1-8(1-\epsilon )\sin (\theta /2)\sin (\varphi /2)\cos ((\theta
-\varphi )/2)+16(1-\epsilon )^{2}\sin ^{2}(\theta /2)\sin ^{2}(\varphi /2)).
\nonumber \\
&&  \label{Dev2}
\end{eqnarray}

See Fig.1 for $D(\theta ,\varphi )$. When $\theta =\varphi $, (\ref{Dev2})
becomes

\begin{eqnarray}
D(\theta ,\theta )=\epsilon (4(1-\epsilon )\sin ^{2}(\theta /2)-1)^{2}.
\label{equalphase1}
\end{eqnarray}

When $\theta =\varphi =\pi /3$ are chosen as phase shifts, the deviation $%
D(\pi /3,\pi /3)=\epsilon ^{3}$\cite{Grover05}.

\subsection{The deviation does not vanish at different phase shifts}

$D(\theta ,\varphi )$ in (\ref{Dev2}) is rewritten as follows.
\begin{eqnarray}
&&D(\theta ,\varphi )=\epsilon \lbrack (4(1-\epsilon )\sin (\theta /2)\sin
(\varphi /2)-\cos ((\theta -\varphi )/2))^{2}+\sin ^{2}((\theta -\varphi
)/2)].  \nonumber \\
&&  \label{Dev4}
\end{eqnarray}%
From (\ref{Dev4}), it is straightforward that $D(\theta ,\varphi )=0$ if and
only if $\theta =\varphi $ and $\cos \theta =1-\frac{1}{2(1-\epsilon )}$,
where $\epsilon \leq 3/4$. It means that the deviation vanishes only when
two phase shifts are equal. Therefore if $\epsilon $ is known and $\epsilon
\leq 3/4$, then we choose $\theta =\arccos (1-\frac{1}{2(1-\epsilon )})$ as
two equal phase shifts to make the deviation vanish. However, as pointed out
before, $\epsilon $ is always close to $1$.

\section{Different phase shifts for small deviation}

From (\ref{Dev2}) and (\ref{equalphase1}) we calculate
\begin{eqnarray}
&&D(\theta ,\varphi )-D(\theta ,\theta )=  \nonumber \\
&&8\epsilon (1-\epsilon )\sin (\theta /2)\sin ((\theta -\varphi
)/2)[\epsilon \cos (\varphi /2)+(1-\epsilon )\cos ((2\theta +\varphi )/2)].
\nonumber \\
&&  \label{Dev3}
\end{eqnarray}%
Let us reduce (\ref{Dev3}) as follows. $\epsilon \cos (\varphi
/2)+(1-\epsilon )\cos ((2\theta +\varphi )/2)$ in (\ref{Dev3})$=$ $\epsilon
(\cos (\varphi /2)-\cos ((2\theta +\varphi )/2))+\cos ((2\theta +\varphi
)/2)=2\epsilon \sin ((\theta +\varphi )/2)\sin (\theta /2)+\cos ((2\theta
+\varphi )/2)$. Then (\ref{Dev3}) is rewritten as follows.

\begin{eqnarray}
&&D(\theta ,\varphi )-D(\theta ,\theta )=  \nonumber  \label{Dev33} \\
&&8\epsilon (1-\epsilon )\sin (\theta /2)\sin ((\theta -\varphi
)/2)[2\epsilon \sin ((\theta +\varphi )/2)\sin (\theta /2)+\cos ((2\theta
+\varphi )/2)].  \nonumber \\
&&  \label{rewritedev3}
\end{eqnarray}

Following (\ref{rewritedev3}), we have the following results.

Result 1.

The deviation for different phase shifts $\theta $ and $\varphi $ is smaller
than the deviation for equal phase shifts $\theta $, i.e., $D(\theta
,\varphi )<D(\theta ,\theta )$, if $0<\theta <\varphi $ and $\epsilon >-\cos
((2\theta +\varphi )/2)/(2\sin ((\theta +\varphi )/2)\sin (\theta /2))$ or $%
0\leq \varphi <\theta $ and $\epsilon <-\cos ((2\theta +\varphi )/2)/(2\sin
((\theta +\varphi )/2)\sin (\theta /2))$.

Result 2.

$D(\theta ,\varphi )<D(\theta ,\theta )$ whenever $0<\theta <\varphi $ and $%
0<(2\theta +\varphi )<\pi .$

The following example follows result 1 immediately.

\noindent Example 1. When $\theta =\pi /3$ and $\pi /3<\varphi <\pi $, $%
D(\pi /3,\varphi )<D(\pi /3,\pi /3)$ for $\epsilon >-\cos (\pi /3+\varphi
/2)/\sin (\pi /6+\varphi /2)$.

It means that the deviation for one phase shift of $\pi /3$ and one larger
phase shift is smaller than for two equal phase shifts of $\pi /3$.

Example 2. $D(\theta ,\pi /2)<D(\theta ,\theta )$ when $0<\theta <\pi /2$
and $\epsilon $ is large.

The proof is as follows.

When $0<\theta \leq \pi /4$, from result 2 $D(\theta ,\pi /2)<D(\theta
,\theta )$ for any $\epsilon $ in $(0,1)$. When $\pi /4<\theta <\pi /2$,
from result 1 $D(\theta ,\pi /2)<D(\theta ,\theta )$ for $\epsilon >(\sin
\theta -\cos \theta )/(2(\sin (\theta /2)+\cos (\theta /2))\sin (\theta /2))$%
.

This example can also be verified by computing $D(\theta ,\pi /2)$ and
reducing $D(\theta ,\pi /2)-D(\theta ,\theta )$ as follows.

$D(\theta ,\pi /2)-D(\theta ,\theta )=4\epsilon (1-\epsilon )\sin (\theta
/2)(\sin (\theta /2)-\cos (\theta /2)+2(1-\epsilon )\sin (\theta /2)\cos
\theta )$. It is not hard to see that $D(\theta ,\pi /2)<D(\theta ,\theta )$
if $\epsilon >1-[(\cos (\theta /2)-\sin (\theta /2))/(2\sin (\theta /2)\cos
\theta )]$.

\section{The phase shifts for the smallest average deviation}

\subsection{The definition of the average deviation}

Usually, people pay much attention to the average-case behavior of an
algorithm besides the worst-case behavior. For Grover's fixed-point search,
Tulsi et al. studied the average-case behavior in \cite{Tulsi}. As indicated
in \cite{Grover98}, $||U_{ts}||$ is very small, i.e., $\epsilon $ is very
large and close to 1. Unfortunately, we don't know the exact value of $%
\epsilon $. However, it may be possible to know an range of $\epsilon $.
From (\ref{Dev2}), it is clear that the deviation is a function of $\epsilon
$. Let $\epsilon $ lie in the range $(\beta ,\alpha )$. Intuitively, the
deviation varies as $\epsilon $ does in $(\beta ,\alpha )$. What is the
average value of the deviation? It seems essential and significant to define
the average deviation in the present letter.

Assume that $\epsilon $ is in the range $(\beta ,\alpha )$, where $0\leq
\beta <\alpha \leq 1$. In terms of mean-value theorem for integrals,\ the
average value $\bar{D}(\theta ,\varphi )$ over the range $(\beta ,\alpha )$
of $\epsilon $\ of deviation $D(\theta ,\varphi )$ is defined and calculated
as follows.

\begin{eqnarray}
&&\bar{D}(\theta ,\varphi )=\frac{1}{\alpha -\beta }\int_{\beta }^{\alpha
}D(\theta ,\varphi )d\epsilon =  \nonumber \\
&&\frac{1}{\alpha -\beta }[\frac{1}{2}(\alpha ^{2}-\beta ^{2})+A\sin (\theta
/2)\sin (\varphi /2)\cos ((\theta -\varphi )/2)+B\sin ^{2}(\theta /2)\sin
^{2}(\varphi /2)],  \nonumber \\
&&  \label{average1}
\end{eqnarray}

\noindent where $A=-\frac{4}{3}(\alpha ^{2}(3-2\alpha )-\beta ^{2}(3-2\beta
))$ and $B=-\frac{4}{3}((1-\alpha )^{3}(3\alpha +1)-(1-\beta )^{3}(3\beta
+1))$.

We argue $A<0$, $B>0$ and $A/B<-1/2$\ in Appendix A.

For example, when $\theta =\varphi =\pi /3$, the average deviation $\bar{D}%
(\pi /3,\pi /3)=$ $(\alpha +\beta )(\alpha ^{2}+\beta ^{2})/4$.

\subsection{The phase shifts for the smallest average deviation}

Apparently, the average deviation $\bar{D}(\theta ,\varphi )$ in (\ref%
{average1})\ is a function of phase shifts $\theta $ and $\varphi $. It is
natural to ask what phase shifts can make the average deviation $\bar{D}%
(\theta ,\varphi )$ the smallest. For this purpose, we need to find the
minimum of the average deviation $\bar{D}(\theta ,\varphi )$. \ \

To find the extremes of the average deviation $\bar{D}(\theta ,\varphi )$,
we compute the following partial derivatives:
\begin{eqnarray}
\partial \allowbreak \bar{D}(\theta ,\varphi )/\partial \theta
&=&1/(2(\alpha -\beta ))\sin (\varphi /2)(A\cos ((2\theta -\varphi
)/2)+B\sin \theta \sin (\varphi /2)),  \nonumber  \label{Dev5} \\
&&
\end{eqnarray}%
\begin{eqnarray}
\partial \allowbreak \bar{D}(\theta ,\varphi )/\partial \varphi
&=&1/(2(\alpha -\beta ))\sin (\theta /2)(A\cos ((2\varphi -\theta )/2)+B\sin
\varphi \sin (\theta /2)).  \nonumber  \label{Dev6} \\
&&
\end{eqnarray}

The extremes of the average deviation $\bar{D}(\theta ,\varphi )$ are $%
(\alpha +\beta )/2$\ at the extreme points of $\bar{D}(\theta ,\varphi )$: $%
\{0,0\},\{0,\pi \}$ and $\{\pi ,0\}$. Let $A\cos ((2\theta -\varphi
)/2)+B\sin \theta \sin (\varphi /2)=0$ in (\ref{Dev5}) and $A\cos ((2\varphi
-\theta )/2)+B\sin \varphi \sin (\theta /2)=0$ in (\ref{Dev6}). Then we
derive the extreme points $\theta =\varphi $. Letting $\theta =\varphi $, $%
\bar{D}(\theta ,\varphi )$ becomes

\begin{eqnarray}
\bar{D}(\theta ,\theta )=((\alpha ^{2}-\beta ^{2})/2+A\sin ^{2}(\theta
/2)+B\sin ^{4}(\theta /2))/(\alpha -\beta ).  \label{equal-pha}
\end{eqnarray}

\noindent After rewriting,

\begin{eqnarray}
\bar{D}(\theta ,\theta )=\frac{B}{4(\alpha -\beta )}(1+\frac{A}{B}-\cos
\theta )^{2}+\frac{\alpha +\beta }{2}-\frac{A^{2}}{4B(\alpha -\beta )}.
\label{Dev7}
\end{eqnarray}

Let us prove that the smallest average deviation occurs at $\theta =\phi $.

\subsubsection{The smallest average deviation for $\protect\alpha +\protect%
\beta \geq 1$}

\noindent For large $\epsilon $,\ the range $(\beta ,\alpha )$ of $\epsilon $
may satisfy $\alpha +\beta \geq 1$. When $\alpha +\beta \geq 1$, $A/B\leq -1$%
. See (3) in Appendix A. In (5) of Appendix A, we demonstrate when $\alpha
+\beta \geq 1$,\ the average deviation $\bar{D}(\theta ,\varphi )\leq
(\alpha +\beta )/2$. It implies that the smallest average deviation occurs
at equal phase shifts. From (\ref{Dev7}), let us find the smallest average
deviation as follows.

Case 1. $-2\leq A/B\leq -1$

When $\theta =\arccos (1+A/B)$ is chosen as equal phase shifts, $\bar{D}%
(\theta ,\theta )$ reaches its minimum $(\alpha +\beta )/2-A^{2}/(4B(\alpha
-\beta ))$, which is also the minimum of $\bar{D}(\theta ,\varphi )$. So, $%
\arccos (1+A/B)$, which is in $[\pi /2$, $\pi ]$, is called the smallest
average deviation point.

Example 3. Let $\beta =0$ and $\alpha =1$. Then $A=-4/3$, $B=4/3$ and $%
1+A/B=0$. $\bar{D}(\theta ,\theta )$ is calculated as $\bar{D}(\theta
,\theta )=\cos ^{2}\theta /3+1/6$. Straightforwardly, $\pi /2$ is the
smallest average deviation point at which $\bar{D}(\theta ,\theta )$ reaches
its minimum $1/6$. See Fig. 2.

Case 2. $A/B<-2$

$\bar{D}(\theta ,\theta )$ decreases as $\theta $ increases from $0$ to $\pi
$ and reaches its minimum $\frac{\alpha +\beta }{2}+4\left( \alpha -1+\beta
\right) \left( \alpha ^{2}-\alpha +\beta ^{2}-\beta \right) $ at $\theta
=\pi $, which is also the minimum of $\bar{D}(\theta ,\phi )$.

Example 4. Let $\beta =3/4$ and $\alpha =1$. Then $A=-5/24$, $B=13/192$, $%
1+A/B=\allowbreak -\frac{27}{13}$. $\bar{D}(\theta ,\theta )$ becomes $%
(13/192)(27/13+\cos \theta )^{2}+73/312$ and $\bar{D}(\theta ,\theta )$
reaches its minimum $\allowbreak \frac{5}{16}$ at $\theta =\pi $. See Fig. 3.

\subsubsection{The smallest average deviation for $\protect\alpha +\protect%
\beta <1$}

When $\alpha +\beta <1$, $-1<A/B<-1/2$. Then it can be shown that when $%
\theta =\arccos (1+A/B)$, $\bar{D}(\theta ,\theta )$ reaches its minimum $%
(\alpha +\beta )/2-A^{2}/(4B(\alpha -\beta ))$, which is less than $(\alpha
+\beta )/2$, and when $\theta =\pi $, $\bar{D}(\theta ,\theta )$ reaches its
maximum $\frac{\alpha +\beta }{2}+4\left( \alpha -1+\beta \right) \left(
\alpha ^{2}-\alpha +\beta ^{2}-\beta \right) $, which is greater than $%
(\alpha +\beta )/2$.

As stated already, the extremes of $\bar{D}(\theta ,\phi )$ at $%
\{0,0\},\{0,\pi \}$ and $\{\pi ,0\}$\ are $(\alpha +\beta )/2$.
Consequently, when $\alpha +\beta <1$, the smallest average deviation occurs
at $\theta =\phi =\arccos (1+A/B)$, which is in $(\pi /3$, $\pi /2)$, and
the maximal average deviation occurs at $\theta =\phi =\pi $.

Remark:

Conclusively, for any range $(\beta ,\alpha )$ of $\epsilon $, $\alpha $ and
$\beta $ determine the phase shifts at which the smallest average deviation
occurs, the smallest average deviation points are greater than $\pi /3$ and
the smallest average deviations are smaller than the average deviation $\bar{%
D}(\pi /3,\pi /3)$\ for equal phase shifts of $\pi /3$.\bigskip

\section{Summary}

In this letter, we demonstrate the possibility of the fixed-point quantum
search algorithm with two different phase shifts. Intuitively, not only
there are more choices for phase shifts to adjust an algorithm for future
physical realization, but also we can find some different phase shifts for
small deviation. Thus, more loose constraint opens a door for more feasible
or robust realization. In this letter, We also show that the smallest
average deviation can be obtained by choosing the following equal phase
shifts. Let $\left( \beta ,\alpha \right) $ be the range of $\epsilon $.
Then if $A/B\geq -2$, then $\arccos (1+A/B)$ is chosen as equal phase
shifts. Otherwise, the closer to $\pi $ the equal phase shifts are, the
better.

Appendix A

1. The proof of $B>0$

We can factor $B$ as follows.

$B=-\frac{4}{3}((1-\alpha )^{3}(3\alpha +1)-(1-\beta )^{3}(3\beta
+1))\allowbreak =$ $\frac{4}{3}\left( \alpha -\beta \right) D$, where $%
D=3\alpha ^{3}-8\alpha ^{2}+3\beta \alpha ^{2}+6\alpha -8\beta \alpha
+3\beta ^{2}\alpha +6\beta -8\beta ^{2}+3\beta ^{3}\allowbreak $. So, we
need only to show that $D>0$. $D$ can be rewritten as $(\alpha +\beta
)[3(\alpha +\beta )^{2}-8(\alpha +\beta )+6]-\alpha \beta \lbrack 6(\alpha
+\beta )-8]$. There are two cases. Case 1: $6(\alpha +\beta )-8\leq 0$.
Clearly $D>0$ since $3(\alpha +\beta )^{2}-8(\alpha +\beta )+6>0$. Case 2: $%
6(\alpha +\beta )-8>0$. Also, $D>0$ since $(\alpha +\beta )[3(\alpha +\beta
)^{2}-8(\alpha +\beta )+6]>(\alpha +\beta )^{2}[6(\alpha +\beta
)-8]/4>\alpha \beta \lbrack 6(\alpha +\beta )-8]$.

The proof was given by Mr. P.Y. Sun.

2. The proof of $A/B<-1/2$

We demonstrate $B>0$ in Appendix A. Hence, it is easy to see $A/B<-1/2$ if
and only if $2A+B<0$. By factoring, $2A+B=\allowbreak \allowbreak \frac{4}{3}%
\left( \alpha -\beta \right) E$, where $E=3\alpha ^{3}-4\alpha ^{2}+3\beta
\alpha ^{2}-4\beta \alpha +3\beta ^{2}\alpha -4\beta ^{2}+3\beta
^{3}\allowbreak $. We only need to argue $E<0$. Letting $\alpha =\beta
+\gamma $, where $0<\gamma \leq 1$,

$E=\allowbreak 12\beta ^{3}+18\beta ^{2}\gamma +12\beta \gamma ^{2}+3\gamma
^{3}-12\beta ^{2}-12\beta \gamma -4\gamma ^{2}$

$=(12\beta ^{3}+12\beta ^{2}\gamma -12\beta ^{2})+(6\beta ^{2}\gamma
+12\beta \gamma ^{2}-12\beta \gamma )+(3\gamma ^{3}-\allowbreak 4\gamma
^{2})=\allowbreak $

$=12\beta ^{2}(\beta +\gamma -1)+6\beta \gamma (\beta +2\gamma -2)+\gamma
^{2}(3\gamma -4)$

$=\allowbreak 12\beta ^{2}(\alpha -1)+6\beta \gamma (\alpha +\gamma
-2)+\gamma ^{2}(3\gamma -4)<0$.

3. The proof of $A/B\leq -1$ when $(\alpha +\beta )\geq 1$

Since $A+B=\allowbreak 4\left( \alpha -1+\beta \right) \left( \alpha -\beta
\right) \left( \alpha ^{2}-\alpha +\beta ^{2}-\beta \right) $, clearly $%
A/B\leq -1$ when $(\alpha +\beta )\geq 1$.

4. The proof of $A<0$

By factoring, we obtain $A=\allowbreak \frac{4}{3}\left( \alpha -\beta
\right) C$, where $C=2(\alpha ^{2}+\alpha \beta +\beta ^{2})-3(\alpha +\beta
)$. By letting $\alpha =\beta +\gamma $, where $0<\gamma <1$, $C=\allowbreak
6\alpha \beta +2\gamma ^{2}-(6\beta +3\gamma )$. \ It is easy to know $C<0$
since $6\alpha \beta <6\beta $ and $2\gamma ^{2}<3\gamma $.

5. The proof of $\bar{D}(\theta ,\varphi )\leq (\alpha +\beta )/2$ when $%
\alpha +\beta \geq 1$

When $\alpha +\beta \geq 1$, we show that $A/B\leq -1$ in (3) of this
appendix. To show $\bar{D}(\theta ,\varphi )\leq (\alpha +\beta )/2$, from (%
\ref{average1}) we only need to show $A\sin (\theta /2)\sin (\varphi /2)\cos
((\theta -\varphi )/2)+B\sin ^{2}(\theta /2)\sin ^{2}(\varphi /2)\leq 0$.
Let us argue this as follows.

$A\sin (\theta /2)\sin (\varphi /2)\cos ((\theta -\varphi )/2)+B\sin
^{2}(\theta /2)\sin ^{2}(\varphi /2)$

$=B\sin (\theta /2)\sin (\varphi /2)\cos ((\theta -\varphi )/2)[\frac{A}{B}+%
\frac{\sin (\theta /2)\sin (\varphi /2)}{\cos ((\theta -\varphi )/2)}]$.

\noindent Next we compute $\frac{A}{B}+\frac{\sin (\theta /2)\sin (\varphi
/2)}{\cos ((\theta -\varphi )/2)}=\frac{A}{B}+\frac{1}{2}-\frac{1}{2}\frac{%
\cos ((\theta +\varphi )/2)}{\cos ((\theta -\varphi )/2)}$. Since $\frac{A}{B%
}\leq -1$, $\frac{A}{B}+\frac{1}{2}-\frac{1}{2}\frac{\cos ((\theta +\varphi
)/2)}{\cos ((\theta -\varphi )/2)}\leq -\frac{1}{2}-\frac{1}{2}\frac{\cos
((\theta +\varphi )/2)}{\cos ((\theta -\varphi )/2)}=-\frac{1}{2}(1+\frac{%
\cos ((\theta +\varphi )/2)}{\cos ((\theta -\varphi )/2)})=-\frac{\cos
(\theta /2)\cos (\varphi /2)}{\cos ((\theta -\varphi )/2)}\leq 0$.

\end{document}